\documentclass[aps,prb,reprint,superscriptaddress]{revtex4-2}

\usepackage{amsmath}
\usepackage{amssymb}
\usepackage{graphicx}
\usepackage{bm}
\usepackage{hyperref}

\newcommand{\RE}{\textit{RE}}
\newcommand{\Pnmm}{\textit{P}4/\textit{nmm}}
\newcommand{\Pmmn}{\textit{Pmmn}}
\newcommand{\cc}{\textit{\textbf{c}}}
\newcommand{\ab}{\textit{\textbf{ab}}}
\newcommand{\qq}{\textit{\textbf{q}}}
\newcommand{\QQ}{\textit{\textbf{Q}}}

\begin{document}

\title{Antiferromagnetic order and lattice response in DyCuAs$_2$}

\author{M. G. Kim}
\email{mgkim@uwm.edu}
\affiliation{
Department of Physics and Astronomy,
University of Wisconsin-Milwaukee,
Milwaukee, Wisconsin 53211, USA
}

\author{J.-W. Kim}
\affiliation{
Advanced Photon Source,
Argonne National Laboratory,
Argonne, Illinois 60439, USA
}

\author{P. Ryan}
\affiliation{
Advanced Photon Source,
Argonne National Laboratory,
Argonne, Illinois 60439, USA
}

\author{D. Evans}
\affiliation{
Department of Physics,
Simon Fraser University,
Burnaby, British Columbia, Canada
}

\author{E. D. Mun}
\affiliation{
Department of Physics,
Simon Fraser University,
Burnaby, British Columbia, Canada
}

\date{\today}

\begin{abstract}
We report high-resolution synchrotron X-ray diffraction and X-ray resonant magnetic scattering (XRMS) studies of the low-temperature crystal and magnetic structures of DyCuAs$_2$, a member of the \RE{}CuAs$_2$ family exhibiting a resistivity minimum above the antiferromagnetic transition temperature. Synchrotron diffraction measurements reveal that DyCuAs$_2$ preserves tetragonal symmetry down to low temperature within the experimental resolution, although pronounced anomalies in both lattice parameters $a$ and $c$ are observed near the antiferromagnetic transition temperature, $T_{\mathrm N}\approx7$~K, indicating strong magnetoelastic coupling. XRMS measurements at the Dy $L_3$ edge establish commensurate antiferromagnetic ordering below $T_{\mathrm N}$ with AFM Bragg peaks at \qq{} = (0, 0, 0.5). Representation analysis and calculations of the AFM Bragg peak intensities identify the magnetic structure as the $\Gamma_{10}$ representation, consisting of in-plane Dy moments stacked along the \cc{} axis in a $++--$ sequence. The magnetic structure is therefore identical to that previously reported for SmCuAs$_2$. Comparison among DyCuAs$_2$, SmCuAs$_2$, and GdCuAs$_2$ suggests that in-plane AFM order and the associated magnetic frustration on the tetragonal lattice are closely connected to the emergence of the resistivity minimum in the \RE{}CuAs$_2$ family. At the same time, the enhanced lattice response and stronger magnetic-field sensitivity observed in DyCuAs$_2$ imply that magnetoelastic and spin-orbit interactions additionally play important roles in determining the robustness of this anomalous transport behavior.
\end{abstract}


\maketitle

\section{Introduction}

The family of \RE{}CuAs$_2$ (\RE{} $=$ rare earth) exhibits rich and complex magnetic and transport behavior. Compounds with \RE{} $=$ Sm, Gd, Tb, and Dy, which are not Kondo systems, exhibit a pronounced resistivity minimum at low temperatures prior to the onset of antiferromagnetic (AFM) ordering~\cite{Sampathkumaran2003,Sengupta2004,Evans2021}. The conventional Kondo mechanism is generally expected to be strongly suppressed in systems possessing large local magnetic moments or strong magnetic anisotropy, as in these compounds. Consequently, the resistivity minimum observed in \RE{}CuAs$_2$ cannot be explained by the conventional exchange interaction between localized $f$ electrons and conduction electrons~\cite{Maple1978}. It has instead been proposed that strong magnetic frustration stabilizing liquid-like spin states may generate a resistivity minimum even in non-Kondo systems~\cite{Wang2016}. Other proposed mechanisms involve spin-dependent scattering processes associated with the underlying long-range magnetic correlations~\cite{Rozenberg2000,Gerber2016,Das2015}, short-range magnetic correlations~\cite{Fritsch2005,Fritsch2006,Mallik1998}, polaronic effects~\cite{Nyhus1997,Chatterjee2004}, and fluctuations~\cite{Matsushita2005,Hasegawa1972,Geldart1975,Kaiser1992,Karmakar2006}.

Understanding the origin of the resistivity minimum in this family requires detailed knowledge of both the crystal and magnetic structures and their evolution at low temperatures. Previous studies on polycrystalline samples established that \RE{}CuAs$_2$ crystallizes in the tetragonal \Pnmm{} structure at room temperature~\cite{Sampathkumaran2003,Sengupta2004,Brylak1995,Mozharivskyj2000,Mozharivskyj2002,Jemetio2002}. In GdCuAs$_2$, slight P substitution was found to induce an orthorhombic distortion into the \Pmmn{} structure~\cite{Mozharivskyj2000,Mozharivskyj2002}. More recently, low-temperature synchrotron studies revealed that stoichiometric GdCuAs$_2$ exhibits an intrinsic weak orthorhombic distortion together with magnetoelastic anomalies near the resistivity minimum~\cite{Balodhi2023}. In contrast, SmCuAs$_2$ retains tetragonal symmetry from room temperature down to below the AFM ordering temperature~\cite{Kim2026}.

The magnetic structures of \RE{}CuAs$_2$ were previously investigated using neutron powder diffraction (NPD)~\cite{Zhao2017}. PrCuAs$_2$, which does not exhibit a resistivity minimum, orders with propagation vector \qq{} $=(0,0,0.5)$ and moments aligned along the \cc{} axis. In contrast, NdCuAs$_2$ and DyCuAs$_2$, both of which exhibit a resistivity minimum, were proposed to possess moments lying within the \ab{} plane with the same propagation vector. TbCuAs$_2$ and HoCuAs$_2$ exhibit more complex incommensurate magnetic ordering. Owing to the intrinsic limitations of powder diffraction, however, the detailed stacking sequence along the \cc{} axis and the precise in-plane moment direction could not be uniquely determined for most compounds.

Recent X-ray resonant magnetic scattering (XRMS) studies on single crystals have begun to resolve these issues. GdCuAs$_2$ was found to order antiferromagnetically below $T_{N_1}\approx10.6$~K with an incommensurate propagation vector \qq{} $=(\delta,0,0.5)$, followed by a lock-in transition at $T_{N_2}\approx6$~K into a commensurate state with \qq{} $=(1/3,0,0.5)$, Gd moments aligned along the \textbf{\textit{b}} axis, and a $++--$ stacking sequence along the \cc{} axis~\cite{Balodhi2023}. SmCuAs$_2$ retains tetragonal symmetry down to base temperature and exhibits commensurate AFM ordering with \qq{} $=(0,0,0.5)$, in-plane moments, and the same $++--$ stacking sequence~\cite{Kim2026}.

In this work, we investigate the low-temperature crystal and magnetic structures of DyCuAs$_2$, which exhibits a resistivity minimum near 15~K and AFM ordering below $T_{\mathrm N}\approx7$~K together with strong easy-plane anisotropy and a metamagnetic transition near 14~kOe for $H\parallel$ \ab{}~\cite{Evans2021}. Using high-resolution synchrotron X-ray diffraction and XRMS at the Dy $L_3$ edge, we determine both the low-temperature lattice symmetry and the microscopic magnetic structure. We find that DyCuAs$_2$ retains tetragonal symmetry within the experimental resolution down to 4.2~K. XRMS measurements reveal commensurate AFM ordering with propagation vector \qq{} $=(0,0,0.5)$, in-plane Dy moments, and a $++--$ stacking sequence along the \cc{} axis. The resulting magnetic structure is identical to that reported for SmCuAs$_2$. Comparison among DyCuAs$_2$, SmCuAs$_2$, and GdCuAs$_2$ suggests that in-plane AFM order and the associated magnetic frustration on the tetragonal lattice are common ingredients underlying the resistivity minimum in this family of compounds.

\section{Experimental Methods}

Single crystals of DyCuAs$_2$ were grown from high-temperature ternary melts~\cite{Evans2021,Balodhi2023}. The phase purity of the growth batch was examined by room-temperature powder X-ray diffraction using a Rigaku MiniFlex diffractometer. The magnetic and electrical properties of the as-grown single crystals were characterized using a Quantum Design Magnetic Property Measurement System and a Quantum Design Physical Property Measurement System.

High-resolution single-crystal X-ray diffraction measurements were performed on a six-circle diffractometer at beamline 6-ID-B of the Advanced Photon Source (APS), which provides access to a wide region of reciprocal space. Measurements were carried out at the Dy $L_3$ absorption edge ($E = 7.790$~keV). The sample was aligned such that the [0, 1, 0] and [0, 0, 1] directions lay within the scattering plane. The single crystal exhibited high crystalline quality with a full width at half maximum of less than $0.02^\circ$, as determined from longitudinal scans of structural Bragg reflections measured at room temperature. Diffraction measurements were performed at 265~K and at the refrigerator base temperature of 4.2~K.

X-ray resonant magnetic scattering (XRMS) measurements were performed on the same beamline using the same sample geometry. The incident X-ray beam was linearly polarized perpendicular to the vertical scattering plane ($\sigma$ polarization), and scattered intensities were recorded using a two-dimensional area detector. In this scattering geometry, contributions from both the $\sigma-\sigma$ and $\sigma-\pi$ channels can be detected simultaneously while preserving high statistics in the Bragg peak intensity. The magnetic reflections appear at forbidden $(0,0,L)$ positions with half-integer $L$, where no structural Bragg peaks are expected. Consequently, the observed scattering intensity at these positions is expected to originate predominantly from magnetic scattering, corresponding primarily to the $\sigma-\pi$ channel. XRMS measurements were performed between 10~K and 4.2~K.

\section{Results and Discussion}

\begin{figure}[t]
    \centering
    \includegraphics[width=\columnwidth]{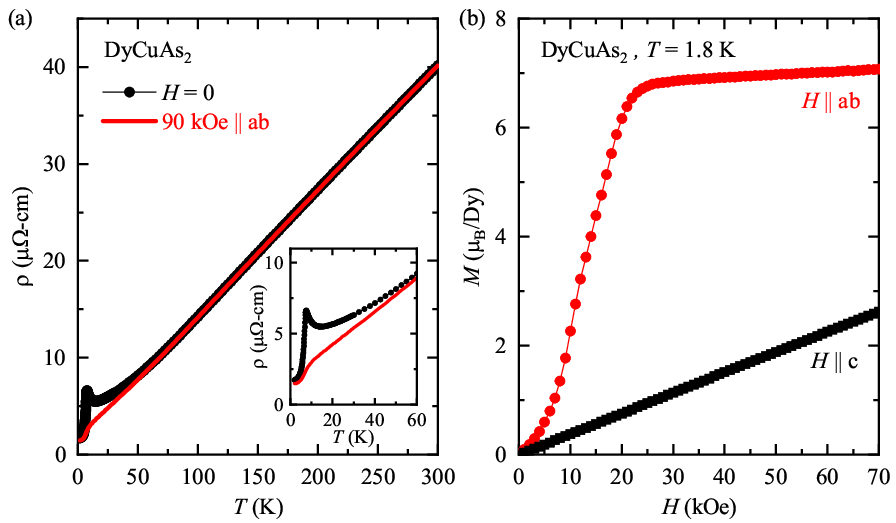}
    \caption{
    (a) Temperature dependence of the electrical resistivity, $\rho(T)$, of DyCuAs$_2$ measured at $H = 0$ (closed symbols) and $H = 90$~kOe (solid line) up to 300~K. Inset: Enlarged view of $\rho(T)$ between 0 and 60~K.
    (b) $M(H)$ curves of DyCuAs$_2$ measured at $T = 1.8$~K for $\mathbf{H}\parallel\mathbf{ab}$ and $\mathbf{H}\parallel\mathbf{c}$.
    }
    \label{fig0}
\end{figure}

Figure~\ref{fig0}(a) presents the electrical resistivity of DyCuAs$_2$. The zero-field electrical resistivity of DyCuAs$_2$ up to 300~K exhibits metallic behavior at high temperatures, followed by an unexpected minimum near 15~K, as shown in Fig.~\ref{fig0}(a). Upon further cooling, $\rho(T)$ increases and reaches a peak at the N\'eel temperature, $T_{\mathrm N}$ ($\approx 7$~K), before decreasing rapidly below the onset of antiferromagnetic ordering. As shown in the inset of Fig.~\ref{fig0}(a), the resistivity minimum is progressively suppressed with increasing magnetic field applied parallel to the $ab$ plane ($\mathbf{H}\parallel\mathbf{ab}$) and disappears completely at $H \geq 30$~kOe.

The anisotropic $M(H)$ curves of DyCuAs$_2$ at 1.8~K are plotted in Fig.~\ref{fig0}(b). For $\mathbf{H}\parallel\mathbf{c}$, the magnetization is linear in field up to 70~kOe. For $\mathbf{H}\parallel\mathbf{ab}$, the magnetization appears to approach saturation near 30~kOe. However, $M(H)$ continues to increase gradually with increasing field, reaching approximately $7~\mu_{\mathrm B}$/Dy at 70~kOe, which remains substantially lower than the theoretical saturation moment of $10~\mu_{\mathrm B}$/Dy for Dy$^{3+}$.

\begin{figure}[t]
    \centering
    \includegraphics[width=\columnwidth]{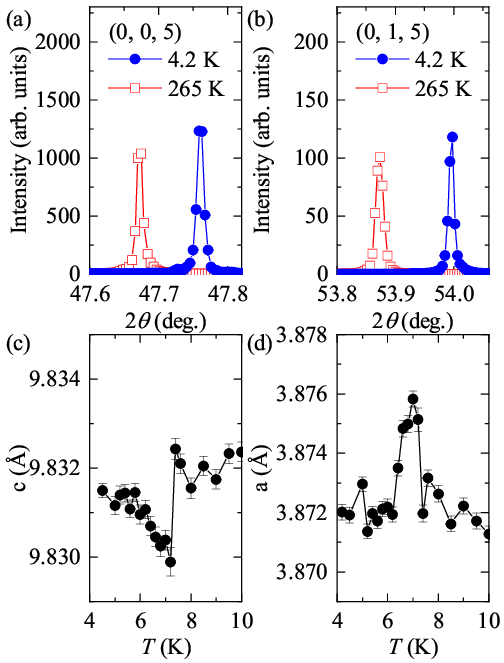}
    \caption{
    (a) Longitudinal ($\theta$--$2\theta$) scans of the $(0,0,5)$ structural Bragg peak measured at 4.2~K and 265~K.
    (b) Longitudinal ($\theta$--$2\theta$) scans of the $(0,1,5)$ structural Bragg peak measured at 4.2~K and 265~K.
    (c) Temperature dependence of the lattice parameter $c$ between 4 and 10~K.
    (d) Temperature dependence of the lattice parameter $a$ between 4 and 10~K.
    }
    \label{fig1}
\end{figure}

Figures~\ref{fig1}(a) and \ref{fig1}(b) show longitudinal ($\theta$--$2\theta$) scans of the structural Bragg peaks $(0,0,5)$ and $(0,1,5)$ measured at 265~K and 4.2~K. Both reflections exhibit a sharp single peak at both temperatures. The observation of single sharp peaks along both the \cc{}-axis direction represented by $(0,0,5)$ and the \textbf{\textit{b}}-axis direction represented by $(0,1,5)$ indicates that the crystal structure remains unchanged between room temperature and low temperature within the experimental resolution. If DyCuAs$_2$ possessed an orthorhombic structure similar to GdCuAs$_2$~\cite{Balodhi2023}, the $(0,1,5)$ reflection would be expected to split into two distinct peaks. Instead, DyCuAs$_2$ exhibits a single peak, similar to SmCuAs$_2$~\cite{Kim2026}, suggesting that the tetragonal structure is preserved down to low temperature.

A closer examination of Fig.~\ref{fig1}(b), however, reveals that the $(0,1,5)$ peak measured at 4.2~K is sharper than that measured at 265~K. The full width at half maximum (FWHM) values for the $(0,0,5)$ reflection at both temperatures and for the $(0,1,5)$ reflection at 4.2~K are approximately $0.011^\circ$, whereas the $(0,1,5)$ peak at 265~K exhibits a broader FWHM of approximately $0.016^\circ$. This broadening may indicate the presence of an unresolved splitting along the \textbf{\textit{b}}-axis direction that remains below the experimental resolution. If this unresolved broadening reflects an orthorhombic splitting below the experimental resolution, the sharpening of the $(0,1,5)$ reflection upon cooling suggests that DyCuAs$_2$ becomes more tetragonal at low temperatures, consistent with the magnetoelastic behavior observed in GdCuAs$_2$. Nevertheless, no direct evidence of peak splitting was resolved in the present measurements.

We additionally measured the $(0,0,5)$ and $(0,1,5)$ structural Bragg peaks between the base temperature (4.2~K) and 10~K. From the measured scattering angle $2\theta$, we extracted the lattice parameter $c$ from the $(0,0,5)$ reflection and the lattice parameter $a$ from the $(0,1,5)$ reflection assuming a tetragonal structure. Figures~\ref{fig1}(c) and \ref{fig1}(d) show the temperature dependence of the lattice parameters. The lattice parameter $c$ exhibits an abrupt decrease near 7~K, followed by a gradual increase upon further cooling. In contrast, the lattice parameter $a$ shows a pronounced peak-like enhancement near 7~K and returns to its higher-temperature value at lower temperatures. Previous bulk measurements identified an AFM transition near 7~K~\cite{Sampathkumaran2003,Sengupta2004,Evans2021}, consistent with the temperature at which anomalies in the lattice parameters are observed.

Such lattice responses accompanying AFM ordering have also been reported in GdCuAs$_2$ and SmCuAs$_2$. In GdCuAs$_2$, the onset of incommensurate magnetic ordering is accompanied by a reduction of the lattice parameter $c$~\cite{Balodhi2023}. However, the magnitude of the lattice anomaly in DyCuAs$_2$ is approximately $0.02\%$, nearly an order of magnitude larger than the $\sim0.002\%$ change observed in GdCuAs$_2$. In contrast, SmCuAs$_2$ exhibits an increase of the lattice parameter $c$ below $T_{\mathrm N}$~\cite{Kim2026}. For the lattice parameter $a$, no pronounced anomaly was observed near $T_{\mathrm N}$ in tetragonal SmCuAs$_2$. In GdCuAs$_2$, the lattice parameter increases by approximately $0.08\%$ near the resistivity-minimum temperature (no significant change at $T_{\mathrm N}$), whereas DyCuAs$_2$ exhibits a comparable increase of approximately $0.1\%$ at the onset of AFM ordering at $T_{\mathrm N}$. Despite occurring at different characteristic temperatures, the similar magnitudes of these lattice responses suggest comparably strong magnetoelastic coupling in DyCuAs$_2$ and GdCuAs$_2$. These pronounced changes in both lattice parameters $a$ and $c$ demonstrate the presence of strong magnetoelastic coupling in DyCuAs$_2$ along both the in-plane and out-of-plane directions.

\begin{figure}[t]
    \centering
    \includegraphics[width=\columnwidth]{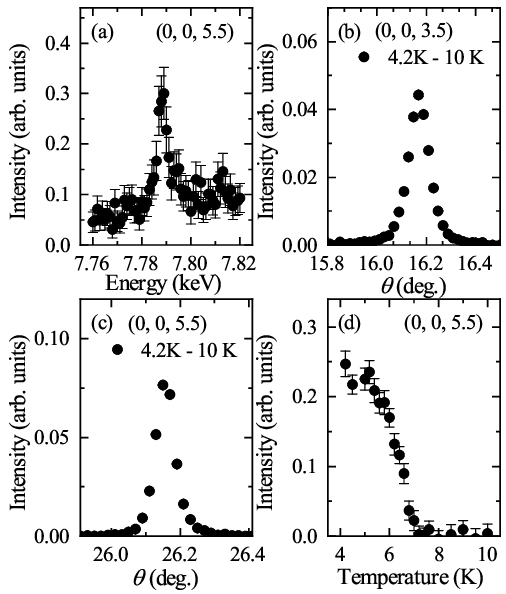}
    \caption{
    (a) Energy scan measured at the $(0,0,5.5)$ AFM Bragg peak position at 4.2~K. The fluorescence background has been subtracted.
    (b) and (c) Longitudinal ($\theta$--$2\theta$) scans measured at 4.2~K and 10~K at the $(0,0,3.5)$ and $(0,0,5.5)$ AFM Bragg peak positions, respectively.
    (d) Temperature dependence of the AFM order parameter obtained at the $(0,0,5.5)$ reflection.
    }
    \label{fig2}
\end{figure}

Figure~\ref{fig2}(a) shows the energy dependence of the scattering intensity at the AFM Bragg peak position $(0,0,5.5)$ at 4.2~K. Because measurements were taken without a polarization analyzer, a strong fluorescence background was initially present; subtracting this background was essential to resolve the peak's resonant behavior. A pronounced resonant enhancement is observed near $E = 7.790$~keV, corresponding to the Dy $L_3$ absorption edge. The observed resonance profile is consistent with an electric dipole ($E1$) transition from the Dy $2p$ core level to the unoccupied $5d$ states~\cite{Detlefs1997,Kim2005,Bouchenoire2009}.

Figures~\ref{fig2}(b) and \ref{fig2}(c) show longitudinal scans measured at the AFM Bragg peak positions $(0,0,3.5)$ and $(0,0,5.5)$, respectively. The data shown correspond to the 4.2~K scans after subtraction of the scans measured at 10~K. Using the two-dimensional detector, we identified a temperature-independent and angle-independent background signal at these AFM Bragg peak positions over the entire measured temperature range, which was subtracted as background. The observed peaks demonstrate that the AFM Bragg reflections occur at commensurate $(0,0,L)$ positions with half-integer $L$. Additional measurements of the $(0,0,2.5)$ and $(0,0,4.5)$ reflections (not shown) further confirmed the commensurate AFM ordering at \qq{} = (0, 0, 0.5).

To investigate the temperature dependence of the magnetic scattering, the $(0,0,5.5)$ reflection was measured between 4.2~K and 10~K. The resulting peaks were fitted using a Lorentzian peak function, and the extracted integrated intensities are shown in Fig.~\ref{fig2}(d). The integrated intensity begins to increase below approximately 7~K, indicating the onset of magnetic scattering at the AFM $(0,0,5.5)$ position. This temperature agrees well with the N\'eel transition temperature previously identified by bulk measurements. These XRMS results therefore establish that DyCuAs$_2$ develops commensurate AFM ordering below $T_{\mathrm N}\approx7$~K.

Previous studies have shown that both PrCuAs$_2$~\cite{Zhao2017} and SmCuAs$_2$~\cite{Kim2026} exhibit AFM ordering with propagation vector \qq{} $=(0,0,0.5)$. PrCuAs$_2$, which does not exhibit a resistivity minimum, was proposed to possess moments aligned along the \cc{} axis, whereas SmCuAs$_2$, which does exhibit a resistivity minimum, was found to possess moments aligned within the \ab{} plane. Since DyCuAs$_2$ also exhibits a resistivity minimum, one may similarly expect the Dy moments to adopt an in-plane alignment analogous to that of SmCuAs$_2$.

Although XRMS can, in principle, determine the magnetic moment direction through the azimuthal angular dependence of the magnetic scattering intensity, for in-plane moments in a tetragonal structure the presence of multiple magnetic domains averages out the angular dependence, making it difficult to distinguish experimentally from the response expected for moments aligned along the \cc{} axis. Therefore, to determine the magnetic structure of DyCuAs$_2$, we measured the \QQ{} dependence of the AFM Bragg peak intensities and compared the results with calculations based on candidate magnetic structure models.

\begin{figure}[t]
    \centering
    \includegraphics[width=\columnwidth]{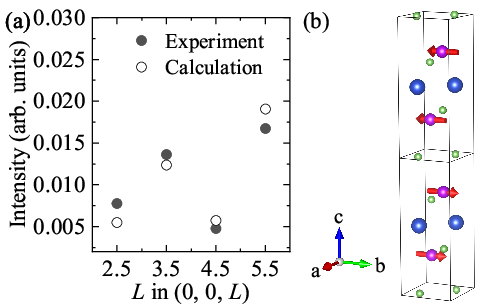}
    \caption{
    (a) Integrated intensities of the AFM Bragg peaks measured at different $L$ values in $(0,0,L)$. Solid symbols represent the experimentally measured intensities, while open symbols denote the calculated intensities for the corresponding AFM Bragg peaks based on the $\Gamma_{10}$ magnetic representation (MR).
    (b) Antiferromagnetic magnetic structure of DyCuAs$_2$ corresponding to the $\Gamma_{10}$ magnetic representation.
    }
    \label{fig3}
\end{figure}

Figure~\ref{fig3}(a) shows the \QQ{} dependence of the AFM $(0,0,L)$ Bragg peaks. The measured intensities reveal that reflections with $L=$ odd $+~0.5$ are significantly stronger than those with $L=$ even $+~0.5$, with the largest intensity observed at $L=5.5$. To identify the magnetic structure responsible for this characteristic \QQ{} dependence, we performed representation analysis~\cite{WILLS2000} for the space group $P4/nmm$ with propagation vector \qq{} $=(0,0,0.5)$. The analysis yields four possible magnetic representations: $\Gamma_2$, $\Gamma_3$, $\Gamma_9$, and $\Gamma_{10}$. The $\Gamma_2$ and $\Gamma_3$ representations correspond to Dy moments aligned along the \cc{} axis, whereas $\Gamma_9$ and $\Gamma_{10}$ correspond to moments aligned within the \ab{} plane. Another important distinction among these representations is the stacking sequence along the \cc{} axis. In $\Gamma_2$ and $\Gamma_9$, the moments adopt a $+--+$ arrangement, whereas in $\Gamma_3$ and $\Gamma_{10}$ they follow a $++--$ stacking sequence.

To compare these candidate magnetic structures with the experimental data, we calculated the AFM $(0,0,L)$ Bragg peak intensities using the relation $I \propto |F|^2$, where the magnetic structure factor is given by
\begin{equation}
F \propto \sum_j f_j^{\mathrm{XRMS}}
e^{i\mathbf{Q}\cdot\mathbf{r}_j}.
\end{equation}
Here, $f^{\mathrm{XRMS}}_{\sigma-\pi}\propto m_a\cos\theta + m_c\sin\theta$, where $m_a$ and $m_c$ denote the magnetic moment components along the $a$ and $c$ directions, respectively. The Lorentz factor was included in the calculated intensities to allow direct comparison with the experimentally measured values. For the $+--+$ stacking sequence realized in $\Gamma_2$ and $\Gamma_9$, the calculated intensities predict stronger peaks at $L=$ even $+~0.5$ than at $L=$ odd $+~0.5$. In contrast, the $++--$ stacking sequence associated with $\Gamma_3$ and $\Gamma_{10}$ produces stronger intensities at $L=$ odd $+~0.5$, consistent with the experimental observations. Consequently, $\Gamma_2$ and $\Gamma_9$ can be excluded as candidate magnetic structures for DyCuAs$_2$.

We further found that the calculated $L$ dependence of the AFM Bragg peak intensities for the $\Gamma_{10}$ representation ($\mathbf{m}\perp$ \cc{}), shown in Fig.~\ref{fig3}(a), reproduces the experimental data well. We can therefore conclude that DyCuAs$_2$ adopts a magnetic structure described by the $\Gamma_{10}$ representation, in which the Dy moments are aligned within the \ab{} plane and stacked along the \cc{} axis in a $++--$ sequence, as illustrated in Fig.~\ref{fig3}(b). This magnetic structure is identical to that previously reported for SmCuAs$_2$~\cite{Kim2026}. Owing to the presence of magnetic domains in the tetragonal structure, however, the precise in-plane orientation of the Dy moments could not be uniquely determined.

The observed magnetic structure may provide important insight into the origin of the resistivity minimum in the \RE{}CuAs$_2$ family. In both DyCuAs$_2$ and SmCuAs$_2$, which exhibit a resistivity minimum, the magnetic moments are aligned within the tetragonal \ab{} plane. Such an in-plane arrangement on the tetragonal lattice can enhance competing magnetic interactions and magnetic frustration, particularly when exchange interactions along symmetry-equivalent in-plane directions cannot be simultaneously satisfied. In contrast, PrCuAs$_2$, which does not exhibit a resistivity minimum, possesses moments aligned along the \cc{} axis. Furthermore, GdCuAs$_2$, which also exhibits a resistivity minimum, develops a more complex magnetic ground state involving incommensurate magnetic ordering with Gd moments aligned along the \textbf{\textit{b}} axis, followed by a lock-in transition into a commensurate phase at lower temperatures. These results suggest that magnetic frustration associated with in-plane AFM order may represent a common underlying ingredient for the emergence of the resistivity minimum in this family of compounds.

Although the magnetic structure of DyCuAs$_2$ is identical to that of SmCuAs$_2$ within the present experimental resolution, the resistivity minimum in DyCuAs$_2$ can be suppressed by a moderate magnetic field ($H = 90$~kOe, see Fig.~\ref{fig0}), whereas the resistivity of SmCuAs$_2$ remains largely insensitive to magnetic fields up to 90~kOe~\cite{Evans2021}. The much stronger magnetic-field dependence of the resistivity minimum in \RE{} = Dy is consistent with its substantially larger magnetic moment ($\approx 10.6\,\mu_{\mathrm B}$) than that of Sm$^{3+}$ ($\approx 0.97\,\mu_{\mathrm B}$)~\cite{Evans2021}, resulting in a correspondingly larger Zeeman energy. Despite the small magnetic moment of Sm$^{3+}$, its relatively large de Gennes factor (4.46) suggests that the exchange field acting on the conduction electrons remains appreciable. In addition, DyCuAs$_2$ exhibits substantially stronger lattice anomalies at $T_{\mathrm N}$ than either SmCuAs$_2$ or GdCuAs$_2$, including pronounced changes in both the lattice parameters $a$ and $c$. These results indicate that DyCuAs$_2$ possesses unusually strong coupling among the lattice, magnetic order, and electronic degrees of freedom. The significantly larger lattice response observed in DyCuAs$_2$ may therefore reflect stronger spin-lattice and spin-orbit coupling effects compared with SmCuAs$_2$, possibly approaching the regime realized in GdCuAs$_2$, where spin-orbit coupling is absent because Gd$^{3+}$ has zero orbital angular momentum. The present results suggest that while the emergence of the resistivity minimum in the \RE{}CuAs$_2$ family is closely connected to in-plane AFM order and the associated magnetic frustration, the robustness of the resistivity minimum against applied magnetic fields is additionally governed by the strength of the magnetoelastic and spin-orbit interactions, which is consistent with the previous discussion in Ref.~\cite{Kim2026}.

\section{Summary}

In summary, high-resolution synchrotron X-ray diffraction and X-ray resonant magnetic scattering measurements establish that DyCuAs$_2$ preserves tetragonal symmetry down to low temperature within the experimental resolution and develops commensurate AFM ordering below $T_{\mathrm N}\approx 7$~K with AFM ordering vector \qq{} $=(0,0,0.5)$. Analysis of the AFM Bragg peak intensities based on representation analysis identifies the magnetic structure as the $\Gamma_{10}$ representation, consisting of in-plane Dy moments stacked along the \cc{} axis in a $++--$ sequence. This magnetic structure is identical to that previously reported for SmCuAs$_2$. In addition, DyCuAs$_2$ exhibits pronounced anomalies in both lattice parameters $a$ and $c$ near $T_{\mathrm N}$, indicating unusually strong magnetoelastic coupling compared with other members of the \RE{}CuAs$_2$ family. Comparison among DyCuAs$_2$, SmCuAs$_2$, and GdCuAs$_2$ suggests that in-plane AFM order and the associated magnetic frustration on the tetragonal lattice are closely connected to the emergence of the resistivity minimum in these compounds. This interpretation is further supported by other \RE{}CuAs$_2$ compounds. In PrCuAs$_2$, where no resistivity minimum is observed, the Pr moments are aligned along the $c$ axis, thereby avoiding magnetic frustration within the $ab$ plane. In contrast, NdCuAs$_2$, which exhibits a resistivity minimum, is known to have magnetic moments lying in the $ab$ plane, likely giving rise to magnetic frustration within the plane. In addition, the stronger field sensitivity of the resistivity minimum and the enhanced lattice response observed in DyCuAs$_2$ imply that magnetoelastic and spin-orbit interactions additionally play important roles in determining the robustness of this anomalous transport behavior.

\begin{acknowledgments}

This work was supported by the University of Wisconsin-Milwaukee.

This research used resources of the Advanced Photon Source, a U.S. Department of Energy (DOE) Office of Science User Facility operated for the DOE Office of Science by Argonne National Laboratory under Contract No. DE-AC02-06CH11357.

E. D. Mun was supported by the Canada Research Chairs, Natural Sciences and Engineering Research Council of Canada, and Canada Foundation for Innovation program.

\end{acknowledgments}

\section*{Author Contributions}

M. G. Kim: Conceptualization, data curation, formal analysis, investigation, methodology, project administration, resources, supervision, validation, visualization, writing---original draft, and writing---review and editing.

J.-W. Kim: Data curation, formal analysis, investigation, methodology, validation, visualization, and writing---review and editing.

P. Ryan: Investigation, methodology, validation, and writing---review and editing.

D. Evans: Investigation and methodology.

E. D. Mun: Investigation, methodology, resources, and writing---review and editing.

\end{document}